\documentclass[showpacs,preprint,showkeys,superscriptaddress,amssymb,nofootinbib,aps]{revtex4-1}

\usepackage{array,multirow}
\usepackage{amssymb}
\usepackage{bm}
\usepackage{mathrsfs}
\usepackage{amsmath}
\usepackage{amsthm}
\usepackage{amssymb}
\def\ni{\noindent}
\def\bee{\begin{equation}}
\def\eee{\end{equation}}
\def\baa{\begin{eqnarray}}
\def\eaa{\end{eqnarray}}

\begin{document}

\title{\Large Revealing hidden symmetries and gauge invariance of the massive Carroll-Field-Jackiw model}

\author{Paulo R. F. Alves}\email{paulo.alves@ice.ufjf.br}
\affiliation{Departamento de F\'{i}sica, Universidade Federal de Juiz de
             Fora--UFJF, 36036-330, Juiz de Fora, MG, Brazil}
\author{Cleber N. Costa}\email{cleber.costa@ice.ufjf.br}
\affiliation{Departamento de F\'{i}sica, Universidade Federal de Juiz de
             Fora--UFJF, 36036-330, Juiz de Fora, MG, Brazil}
         \author{Everton M.\ C.\ Abreu}\email{evertonabreu@ufrrj.br}
         \affiliation{Departamento de F\'{i}sica, Universidade Federal Rural do
         	Rio de Janeiro--UFRRJ, 23890-971, Serop\'edica, RJ,  Brazil}
         \affiliation{Departamento de F\'{i}sica, Universidade Federal de Juiz de
         	Fora--UFJF, 36036-330, Juiz de Fora, MG, Brazil}
         \affiliation{Programa de P\'os-Gradua\c{c}\~ao Interdisciplinar em
         	F\'{\i}sica Aplicada, Instituto de F\'{\i}sica, Universidade
         	Federal do Rio de Janeiro-UFRJ, 21941-972, Rio de Janeiro, RJ,
         	Brazil}
\author{Jorge Ananias Neto}\email{jorge@fisica.ufjf.br}
\affiliation{Departamento de F\'{i}sica, Universidade Federal de Juiz de
             Fora--UFJF, 36036-330, Juiz de Fora, MG, Brazil}
\author{Albert C. R. Mendes}\email{albert@fisica.ufjf.br}
 \affiliation{Departamento de F\'{i}sica, Universidade Federal de Juiz de
         	Fora--UFJF, 36036-330, Juiz de Fora, MG, Brazil}


\begin{abstract}
\noindent
In this paper we have analyzed the improved version of the Gauge Unfixing (GU) formalism of
the massive Carroll-Field-Jackiw model, which breaks both the Lorentz and gauge 
invariances, to disclose hidden symmetries to obtain gauge invariance, the key stone
of the Standard Model.
In this process, as usual, we have converted this second-class system into a first-class one and 
we have obtained two gauge invariant models. 
We have verified that the Poisson brackets involving the gauge invariant
variables, obtained through the GU formalism,  coincide with the Dirac brackets 
between the original second-class variables of the phase space. Finally, we have obtained
two gauge invariant Lagrangians where one of them represents the Stueckelberg form.
\end{abstract}

\pacs{11.15.-q; 11.10.Ef; 11.30.Cp}
\keywords{constrained system, gauge unfixing formalism, Maxwell-Carroll-Field-Jackiw model}

\maketitle

\section{Introduction}

The Caroll-Field-Jackiw (CFJ) model \cite{cfj} is a variation of Maxwell's 
electrodynamics (ME) where, due to the presence of the CFJ term in the 
original Maxwell Lagrangian, the Lorentz invariance is broken. 
The CFJ term contains a Lorentz violating (LV) fixed 
background, $V_{\mu}$, which is possibly responsible for the phenomenon of 
birefringence \cite{ym}. This model is also known as the gauge sector of the 
extended standard CPT-odd model \cite{casana2008lorentz,
alves2016aspectos}. In the language of constrained systems, via Dirac formalism, both ME and CFJ
models are gauge invariant systems with two first-class constraints.
On the other hand, if we add a photon mass term in the original Maxwell Lagrangian,
we have the Abelian Proca model which is not gauge invariant with two second-class 
constraints. Combining the CFJ theory with the Abelian Proca model we have a model 
where both Lorentz invariance and gauge invariance are broken. So, it is possible to disclose 
gauge symmetries in this new model by converting second-class system into a first-class
one. In this context, one formalism which we can mention is the Batalin-Fradkin-Tuytin (BFT) method \cite{bft} 
which is formulated in an extended phase space with the introduction of the Wess-Zumino terms. The other one  
is the Gauge Unfixing (GU)
formalism where gauge theories are obtained within the phase space of
the original second-class theory. The GU formalism was originally formulated by Mitra
and Rajaraman \cite{mr} and continued by Anishetty and Vytheeswaran \cite{vyt,vyt2,dayi,biz}. Since no extra 
variables in the phase space are not used, this procedure seems to be very attractive.

The purpose of this paper is to disclose the hidden gauge invariance and to discuss its consequences concerning the massive CFJ model.  We have disclosed it
through the improved GU formalism \cite{jan,mm} in order
to analyze the massive CFJ models which are not gauge invariant, obviously. We organized our ideas in the
following manner. In section II we described a Hamiltonian formulation of the massive CFJ. In section III we presented a revision of 
an improved GU techniques and an example is carried out. In section IV we applied our method to the massive CFJ model where two gauge theories were derived.  Section V is  devoted to the conclusions and final words.

\section{The canonical structure of the massive CFJ model}

The massive CFJ model is described by the following Lagrangian density

\begin{equation}
\label{ol}
\mathscr{L}=-\frac{1}{4}F{}_{\alpha\nu}F{}^{\alpha\nu}
-\frac{1}{4}\varepsilon{}_{\beta\alpha\rho\varphi}V{}^{\beta}A{}^{\alpha}F{}^{\rho\varphi}+m^{2}A_{\alpha}A^{\alpha} \,,
\end{equation}

\ni where $\varepsilon{}_{\beta\alpha\rho\varphi}V{}^{\beta}A{}^{\alpha}F{}^{\rho\varphi}$ is the term responsible by the broken of the Lorentz symmetry, $m$ is the 
mass of the photon and $g_{\mu\nu}=diag(+---)$. The canonical momenta is given by
\begin{equation}
\label{cm}
\pi^{\mu}=-F^{0\mu}-\frac{1}{2}\epsilon^{0\mu\alpha\beta}V_{\alpha}A_{\beta} \,.
\end{equation}

\ni From Eq.(\ref{cm}) we have the primary constraint
\begin{eqnarray}
\label{pri}
\phi_1\equiv\pi^{0}\thickapprox 0 \,.
\end{eqnarray}

\ni In the phase space we have the fundamental Poisson bracket 
\begin{equation}
\Big\{ A_{\mu}(x),\pi^{\nu}(y)\Big\} =\delta_{\mu}^{\nu}\delta^{3}(x-y)\,\,.
\end{equation}

\ni The momenta $\pi^{k}$ is given by
\begin{equation}
\pi^{k}=\dot{A}^{k}-\partial^{k}A^{0}-\frac{1}{2}\epsilon^{0kij}V_{i}A_{j} \,\,.
\end{equation}

\ni After the Legendre transformation we can write the canonical Hamiltonian as being
\begin{align}
\label{hc}
H_{c}=&\int d^{3}x\bigg\{ \frac{1}{2} \pi_i\pi_i+A_{0}\partial_i\pi_i+\frac{1}{4}F_{ij}F_{ij}\nonumber\\
&+\frac{1}{2}\epsilon_{0kij}\pi_{k}V_{i}A_{j}
+\frac{1}{8}\Big[\epsilon_{0kij}V_{i}A_{j}\Big]^{2}-\frac{1}{4}\epsilon_{0kij}V_{0}A_{k}F_{ij}\nonumber\\
&+\frac{1}{4}\epsilon_{0kij}V_{k}A_{0}F_{ij}-\frac{m^2}{2}(A_{0}^{2}-A^{2}_i)\bigg\} \,\,.
\end{align}

\ni From the time stability condition of the constraint in Eq. (\ref{pri}), we can obtain the secondary constraint
\begin{equation}
\label{sec}
\dot{\phi}_1\,=\,\phi_2\equiv-\partial_{k}\pi_{k}-\frac{1}{4}\epsilon_{0kij}V_{k}F_{ij}+m^2 A_{0}\thickapprox 0\,\,.
\end{equation}

\ni We can observe that no further constraints will be generated via this iterative procedure and $\phi_1$ and $\phi_2$ are the final group constraints of the model. Calculating the Poisson brackets of these constraints, we have
\begin{equation}
\label{fi12}
\Big\{ \phi_1(x),\phi_2(y)\Big\} = -m^2\delta^3(x-y)\,\,,
\end{equation}

\ni and Eq. (\ref{fi12}) shows that the constraints $\phi_1$ and $\phi_2$ have a second-class structure via Dirac constrained systems classification.

\section{The improved GU formalism}

Consider a Hamiltonian system with two second-class constraints, $T_1$
and $T_2$. The basic idea of the GU formalism consists of selecting one of
the two second-class constraints to be the gauge symmetry generator and
the other one will be discarded  as a gauge generator. For example, if we choose $T_1$ as the gauge 
symmetry generator then $T_2$ will be discarded as a gauge generator, obviously. The constraint $T_1$ will
be redefined as $T_1/\Delta_{12}\equiv \widetilde{T}$ where $\Delta_{12}\equiv 
\{T_1,T_2\}$. The Poisson bracket between $\widetilde{T}$ and $T_2$ is $\{\widetilde{T},
T_2\}=1$, so that $\widetilde{T}$ and $T_2$ are canonically conjugate. The second-class 
Hamiltonian needs to be modified in order to satisfy a first-class algebra
and, consequently, to give rise to a first-class Hamiltonian. We can 
build a gauge invariant Hamiltonian by a series in powers of $T_2$ as
\begin{eqnarray}
\label{gh}
\tilde{H}=H+\{H,\tilde{T}\}  T_2+\frac{1}{2!} \{ \{H,\tilde{T}\}, \tilde{T}\}  T_2^2
\nonumber\\
+\frac{1}{3!} \{ \{ \{H,\tilde{T}\}, \tilde{T}\}, \tilde{T}\}  T_2^3
+\dots \,,
\end{eqnarray}

\ni  where, by construction, we can show that $\{\tilde{H}, \tilde{T}\}=0$.

 The improved GU formalism, introduced by one of us \cite{jan,jan2}, modifies the original phase space variables in order to obtain a gauge invariant phase space. Consider the initial phase space variable written in the following form
\begin{equation}
F=F(q_i, p_i)\,\,.
\end{equation}

\ni The gauge invariant variable $\widetilde{F}$ to be obtained must satisfy the variational condition
\begin{equation}
\label{vc}
\delta\widetilde{F}=\epsilon \{\widetilde{F},\widetilde{T}\}=0 \,\,,
\end{equation}

\ni where $\widetilde{T}$ is the constructed second-class constraint that was chosen to be the gauge symmetry generator and $\epsilon$ is an infinitesimal
parameter. Any function of $\widetilde{F}$ will be gauge invariant
since 
\begin{equation}
\left\{ \widetilde{G}(\widetilde{F}),\widetilde{T}\right\} =\left\{ \widetilde{F},\widetilde{T}\right\} \frac{\partial\widetilde{G}}{\partial\widetilde{F}}=0 \,,
\end{equation}

\ni where 
\begin{equation}
\left\{ \tilde{F},\tilde{T}\right\} \frac{\partial\tilde{G}}{\partial\tilde{F}}\equiv\left\{ \tilde{q}_i,\tilde{T}\right\} \frac{\partial\tilde{G}}{\partial\tilde{q}_i}\,+\,\left\{ \tilde{p}_i,\tilde{T}\right\} \frac{\partial\tilde{G}}{\partial\tilde{p}_i}\,\,.
\end{equation}

\ni Consequently, we can obtain a gauge invariant function from the following substitution 
\begin{equation}
G(F)\Rightarrow G(\widetilde{F})=\widetilde{G}(\widetilde{F})\,\,.
\end{equation}


\ni {\bf An example: the Abelian Pure Chern Simons Theory}\\

We will follow \cite{jan} to exemplify the technique.  The CS theory, being a (2+1) dimensional field theory, is governed by the Lagrangian 
\begin{equation}
\label{initial}
L = \int d^2x\,\frac {k} {2}  \,\epsilon^{\mu\nu\rho}\, A_\mu\partial_\nu A_\rho,
\end{equation}

\ni where $\,k\,$ is a constant. From  Dirac's constrained method \cite{Dirac} the three canonical momenta, i.e., the primary constraints are
\begin{eqnarray}
\label{pii}
T_0 \equiv \pi_0\approx 0\,\,, \qquad \qquad
T_i \equiv \pi_i-\frac{k}{2}\, \epsilon_{ij} A^j\approx 0 \;\;(i=1,2).
\end{eqnarray}

\ni Using the Legendre transformation, the canonical Hamiltonian is 
\begin{equation}
\label{hc}
H_c= -k \int d^2 x \,A_0\,\epsilon^{ij}\,\partial_iA_j\,\,.
\end{equation}

\ni and the secondary constraint is
\begin{equation}
\label{pi3}
T_3\equiv k \, \epsilon^{ij}\,\partial_i A_j\approx 0\,\,,
\end{equation}

\ni and no extra constraints are generated.
To split both the second and the first class constraints, we need to redefine the constraint (\ref{pi3}). 
\begin{eqnarray}
\label{bomega3}
\tilde{T}_3\equiv T_3+\partial^i T_i
=\partial^i\pi_i+\frac{k}{2}\,\epsilon^{ij}\,\partial_iA_j.
\end{eqnarray}

\ni Hence, $T_0$ and $\tilde{T}_3$ are first class constraints.  And
 $T_i$, the second one of Eqs. (\ref{pii}), that are second class constraints, which obey the algebra
$\{T_i(x),T_j(y)\}=-k\,\epsilon_{ij}\,\delta^3(x-y) \;\; (i,j=1,2)$.

Let us choose the symmetry gauge generator as
\begin{equation}
\label{g1}
\tilde{T}=-\frac{T_1}{k}=-\frac{\pi_1}{k}+\frac{A_2}{2}.
\end{equation}

\ni Then, we have the algebra $\{\tilde{T}(x),T_2(y)\}=\delta^3(x-y)$. The second class constraint $T_2=\pi_2+\frac{k}{2} A_1$ will be discarded. The gauge transformations generated by symmetry generator $\tilde{T}$ are
\begin{eqnarray}
\label{da}
\delta A_i=\epsilon \{A_i(x),\tilde{T}(y)\}=
-\frac{\epsilon}{k}\,\delta^i_1\,\delta^3(x-y),\\
\label{dpi}
\delta \pi_i=\epsilon \{\pi_i(x),\tilde{T}(y)\}=
-\frac{\epsilon}{2}\,\delta_i^2\,\delta^3(x-y),\\
\delta T_2=\epsilon\, \{T_2(x),\tilde{T}(y)\}=-\epsilon \,\delta^3(x-y).
\end{eqnarray}

\ni The gauge invariant field $\tilde{A}_i$ is constructed by the expansion of $T_2$, i.e.,  
$\tilde{A}_i=A_i+ b_1\,T_2+b_2\,T_2^2+\ldots+b_n\,T_2^n$.
From the invariance condition $\delta\tilde{A}_i=0$, we can calculate all the correction terms $b_n$. For the linear correction term in order of $T_2$, we can write
\begin{eqnarray}
\delta A_i+b_1\delta T_2=0 \;\Rightarrow\; -\frac{\epsilon}{k}\,\delta^i_1\delta^3(x-y)-b_1\epsilon\,\delta^3(x-y)=0 \;\Rightarrow\; b_1=-\frac{1}{k}\,\delta^i_1.
\end{eqnarray}

\ni For the quadratic term, we obtain $b_2=0$, since $\delta b_1=\epsilon\{b_1,\tilde{T}\}=0.$ Hence, all the correction terms $b_n$ with $n\geq 2$ are zero. Therefore, the gauge invariant field $\tilde{A}_\mu$ is
\begin{eqnarray}
\label{ai}
\tilde{A}_0=A_0\,\,, \qquad \qquad 
\tilde{A}_i=A_i-\frac{1}{k}\,\delta^i_1\,T_2,
\end{eqnarray}
or
\begin{eqnarray}
\label{A0}
\tilde{A}_0&=&A_0,\\
\label{A1}
\tilde{A}_1&=&A_1-\frac{1}{k}\,T_2,\\
\label{A2}
\tilde{A}_2&=&A_2,
\end{eqnarray}

\ni where by using Eq.(\ref{da}), it is easy to show that $\delta\tilde{A}_\mu=0$. The gauge invariant field $\tilde{\pi}_i$ is also constructed by the series in powers of $T_2$ 
\begin{equation}
\tilde{\pi}_i=\pi_i+ c_1\,T_2+c_2\,T_2^2+\ldots+c_n\,T_2^n.
\end{equation}

\ni From the invariance condition $\delta\tilde{\pi}_i=0$, we can calculate all the correction terms $c_n$. For the linear correction term in order of $T_2$, we have
\begin{eqnarray}
\delta \pi_i+c_1 \delta T_2=0 \;\Rightarrow\; -\frac{\epsilon}{2}\delta_i^2\delta^3(x-y)-c_1\epsilon\,\delta^3(x-y)=0 \,\Rightarrow\,  c_1=-\frac{1}{2}\delta_i^2.
\end{eqnarray}

\ni For the quadratic term, we obtain that $c_2=0$, since $\delta c_1=\epsilon\{c_1,\tilde{T}\}=0.$ Consequently, all the correction terms $c_n$ with $n\geq 2$ are zero. Therefore, the gauge invariant field $\tilde{\pi}_i$ is
$\tilde{\pi}_i=\pi_i-\frac{1}{2}\,\delta_i^2\,T_2$,
or
$\tilde{\pi}_1=\pi_1$ and 
$\tilde{\pi}_2=\pi_2-\frac{1}{2}\,T_2$,
where, by using Eq.(\ref{dpi}), it can be shown that $\delta\tilde{\pi}_i=0$. The Poisson brackets between the gauge invariant fields are
\begin{eqnarray}
\label{aiaj}
\{\tilde{A}^i(x),\tilde{A}^j(y)\}=\frac {1}{k}\,\epsilon^{ij}\,\delta^3(x-y),\\
\label{pipj}
\{\tilde{\pi}_i(x),\tilde{\pi}_j(y)\}=\frac {k}{4}\,\epsilon_{ij}\,\delta^3(x-y),\\
\label{aipij}
\{\tilde{A}^i(x),\tilde{\pi}_j(y)\}=\frac {1}{2}\,\delta^i_j\,\,\delta^3(x-y).
\end{eqnarray}

\ni We can observe that the Poisson brackets, Eqs. (\ref{aiaj}), (\ref{pipj}) and (\ref{aipij}), can be written as the original Dirac brackets \cite{pcs} since $T_2=0$. The gauge invariant Hamiltonian, written only in terms of the original phase space variables, is obtained by substituting $A_\mu$ by $\tilde{A}^\mu$, Eqs. (\ref{ai}) and (\ref{A0}), in the canonical Hamiltonian, Eq. (\ref{hc}), such that
\begin{eqnarray}
\label{fh}
\tilde{H}=k \int d^2x\ \,\epsilon^{ij}\partial_i\tilde{A}^0\,\,\tilde{A}_j=
H_c+\int d^2x \;\partial^2A_0\;T_2\nonumber\\ \nonumber\\
=\int d^2x\;[k\, \epsilon^{ij} \partial_i A_0\, \, A_j + \partial^2 A_0\,\pi_2+\frac{k}{2}\,\partial^2 A_0\, A_1].
\end{eqnarray}

\ni We can use the stability condition of $\pi_0\,(T_0\equiv\pi_0)$
\begin{eqnarray}
\label{hpi0}
\{\pi_0,\tilde{H}\}=0 \; &\Rightarrow& \; k \, \epsilon^{ij}\,\partial_i A_j+\partial^2\pi_2+\frac{k}{2}\partial^2A_1= k \, \epsilon^{ij}\,\partial_i A_j+\partial^2 T_2=0 \nonumber\\ \nonumber\\ 
&\Rightarrow& k \; \epsilon^{ij}\,\partial_i \tilde{A}_j=0,
\end{eqnarray}

\ni we have the secondary constraint
\begin{equation}
\tilde{T}_3\equiv k \, \epsilon^{ij}\,\partial_i \tilde{A}_j\,\,,
\end{equation}

\ni which is just the secondary constraint, Eq.(\ref{pi3}), with the substitution of $A_i$ by $\tilde{A}^i$. The gauge invariant Hamiltonian $\tilde{H}$ and the irreducible constraints $T_0,\tilde{T} $ and $\tilde{T}_3$ form a set of first class algebra given by
\begin{eqnarray}
\{\tilde{H},\tilde{T}\}=0,\\
\{\tilde{H},T_0\}=\tilde{T}_3,\\
\label{ht3}
\{\tilde{H},\tilde{T}_3\}=0,\\
\label{tt3}
\{\tilde{T},\tilde{T}_3\}=0,\\
\{\tilde{T},T_0\}=0,\\
\{T_0,\tilde{T}_3\}=0,
\end{eqnarray}

\ni where we have used relation (\ref{aiaj}) to prove Eq.(\ref{ht3}) and the condition $\,\delta \tilde{A}^i=0\,$ to prove Eq.(\ref{tt3}). 
To sum up, firstly, by imposing the stability of $T_0$, Eq.(\ref{hpi0}), we obtain, by a systematic way, an irreducible first class constraint $\tilde{T}_3$. Secondly, we only embed  the initial second class constraint $T_1$, Eq.(\ref{pii}), and, consequently, we have all the constraints that form a first class set.  Besides, in order to reduce all the constraints of the CS theory in a second class nature it is enough to assume $T_2=0$. 

Finally, the gauge invariant CS Lagrangian can be deduced by performing the inverse Legendre transformation
$\tilde{L}=\int d^2x\; (\,\tilde{\pi}_i\dot{\tilde{A}^i}-\tilde{H}\,)$,
where $\tilde{H}$ is given by Eq.(\ref{fh}). As the gauge invariant Hamiltonian, $\tilde{H}$, has the same functional form of the canonical Hamiltonian, Eq.(\ref{hc}), thus, from the inverse Legendre transformation just above we can deduce that the first class Lagrangian (written in terms of the first class variables) will take the same functional form of the original Lagrangian, Eq. (15)
$\tilde{L}=\int d^2x\,\frac {k} {2}  \,\epsilon^{\mu\nu\rho}\, \tilde{A}_\mu\partial_\nu \tilde{A}_\rho$.
Using Eqs.(\ref{A0}), (\ref{A1}) and (\ref{A2}), the gauge invariant Lagrangian, Eq.(15), becomes
\begin{eqnarray}
\label{fl22}
\tilde{L}=\int d^2x\,\frac {k} {2}\, [\, A_0\partial^1A_2-A_0\partial^2 A_1+\frac{1}{k}\,A_0\partial^2 T_2\nonumber\\+A_1\partial^2A_0-\frac {1}{k}T_2\,\partial^2A_0-A_1\partial^0A_2+\frac{1}{k}T_2\,\partial^0A_2\nonumber\\
+A_2\partial^0A_1- \frac{1}{k}A_2\partial^0T_2-A_2\partial^1A_0\,]\,.
\end{eqnarray}

\ni The Hamiltonian equation of motion provides a relation for $\partial^0 A_2$ given by
\begin{equation}
\label{a2}
\partial^0 A_2=\{A_2,\tilde{H}\}=\partial^2 A_0.
\end{equation}

\ni Then, using Eq.(\ref{a2}) and after some algebra using Eq.(\ref{fl22}), we obtain
\begin{eqnarray}
\label{gl}
\tilde{L}&=&\int d^2x\,\frac{k}{2}\, \Big[ A_0\partial^1 A_2 - A_0\partial^2 A_1 + A_1 \partial^2 A_0-A_1\partial^0 A_2 
+A_2 \partial^0 A_1 - A_2\partial^1 A_0 \Big] \nonumber \\
&=&\int d^2x\,\frac {k} {2}  \,\epsilon_{\mu\nu\rho}\, A_\mu\partial^\nu A_\rho.
\end{eqnarray}

\ni We can observe that the gauge invariant Lagrangian, Eq.(\ref{gl}), reduces to the original Lagrangian, Eq.(\ref{initial}). 
The relation (\ref{gl}) is also an important result because without the presence of the extra terms in the gauge invariant Lagrangian, the original gauge symmetry transformation $\;A_\mu\rightarrow A_\mu+\partial_\mu \Lambda \;$ is certainly maintained. 


\section{Raising gauge theories in the massive CFJ model}

We will now compute the underlying symmetries of the massive CFJ model by using the improved GU formalism \cite{jan,jan2}. From the two second-class constraints, Eqs. (\ref{pri}) and (\ref{sec}), we have two possible choices for the gauge symmetry generator. We consider these ones separately.


\ni Case (i) ($\phi_1$ is the gauge symmetry generator)

 
\ni  We will begin by redefining the constraint $\phi_1$, Eq. (\ref{pri}), which was initially chosen to be the symmetry gauge generator, as
\begin{equation}
\tilde{\phi}=-\,\frac{\phi_1}{m^2} \,,
\end{equation}

\ni so that 
\begin{equation}
\left\{ \phi_2(x),\tilde{\phi}(y)\right\} =-\,\delta^3(x-y)\,\,.
\end{equation}

\ni The gauge invariant variable $A_{0}$ is constructed by the power series of the integral in $\phi_2$
\begin{equation}
\label{gip}
\tilde{A_0}(x)=A_0(x)
+\int d^{3}y\,C_{1}(x,y)\,\phi_2(y)+\int\int d^{3}y\,d^{3}z\,C_{2}(x,y,z)\,\phi_2(y)\,\phi_2(z)+... \,\,.
\end{equation}

\ni The coefficients $\,C_{n}\,$ in Eq. (\ref{gip}) are then determined by the variational condition 
\begin{eqnarray}
\label{da0}
\delta\tilde{A}_0(x)=\delta A_0(x)+\int d^{3}y\delta C_{1}(x,y)\phi_2(y)
+\int d^{3}yC_{1}(x,y)\delta \phi_2(y)\nonumber \\+\int\int d^{3}yd^{3}z\delta C_{2}(x,y,z)\phi_2(y)\phi_1(z)
 +\int\int d^{3}yd^{3}zC_{2}(x,y,z)\delta \phi_2(y)\phi_1(z)\nonumber \\+\int\int d^{3}yd^{3}zC_{2}(x,y,z) \phi_2(y)\delta \phi_2(z)+...=0.
\end{eqnarray}

\ni From Eq. (\ref{da0}) we can derive the zeroth order equation in $\phi_2$ 
\begin{equation}
\label{d0}
\delta A_0(x)+\int d^{3}yC_{1}(x,y)\delta \phi_2(y)=0 \,.
\end{equation}

\ni Using 
\begin{eqnarray}
\label{da02}
\delta A_0(x)=\epsilon \{A_0(x), \tilde{\phi}(y)\} 
=-\frac{\epsilon\,\delta^{3}(x-y)}{m^2}\,,
\end{eqnarray}

\ni and 
\begin{eqnarray}
\label{dfi}
\delta \phi_2(y)=\epsilon\left\{ \phi_2(y),\tilde{\phi}(x)\right\} =-\epsilon\,\delta^{3}(x-y) \,,
\end{eqnarray}

\ni in Eq. (\ref{d0}) we find that
\begin{equation}
\label{c1}
C_{1}(x)=-\frac{\delta^{3}(x-y)}{m^2}\,.
\end{equation}

\ni The linear equation in $\phi_2$ is
\begin{align}
\label{d1}
&\int d^{3}y\delta C_{1}(x,y)\phi_2(y)+\int\int d^{3}yd^{3}zC_{2}(x,y,z)\delta \phi_2(y)\phi_2(z)\nonumber\\
&+\,\int\int d^{3}yd^{3}zC_{2}(x,y,z)\phi_2(y)\delta \phi_2(z)=0 \,.
\end{align}

\ni Using Eqs. (\ref{dfi}) and (\ref{c1}) in Eq. (\ref{d1}) we can obtain the coefficient $C_{2}(x,y,z)$ 
\begin{equation}
\label{c2}
C_{2}(x,y,z)=\frac{\delta C_{1}(x)}{2\epsilon}=-\frac{\{\delta^{3}(x-y),\tilde{\phi}\}}{2m^2}=0\,.
\end{equation}
 
\ni Since $C_{2}(x,y,z)=0$ then all the correction terms $C_n$ with $ n \geq 2$ are zero.
Using Eq.(\ref{c1}) into Eq. (\ref{gip}) we obtain the gauge invariant field $\tilde{A_{0}}$ given by
\begin{eqnarray}
\label{gv}
\tilde{A_{0}}=A_0-\frac{\phi_2}{m^2} 
=\frac{1}{m^2}\left\{ \partial_{k}\pi_{k}+\frac{1}{4}\epsilon_{0kij}V_{k}F_{ij}\right\} \,.
\end{eqnarray}

\ni The others fields $A_{i}$, $\pi_{0}$, $\pi_{i}$, are not modified
due to relation
\begin{equation}
\delta\pi_{0}=\delta A_{i}=\delta\pi_{i}=0 \,.
\end{equation}

\ni Therefore  we have that
\begin{align}
\label{gv12}
\tilde{\pi}_{0}&=\pi_{0}\,, \\ 
\label{gv13}
\tilde{A_{i}}&=A_{i}\,, \\
\label{gv14}
\tilde{\pi}_{i}&=\pi_{i}.
\end{align}

\ni The Poisson brackets between the gauge invariant fields are
\begin{align}
\label{pb1}
\left\{ \tilde{A_{0}}(x),\tilde{A_{0}}(y)\right\} &=\left\{ \tilde{\pi_{0}}(x),\tilde{\pi_{0}}(y)\right\} =\left\{ \tilde{A_{0}}(x),\tilde{\pi}_{0}(y)\right\} \nonumber\\
=\left\{ A_{0}(x),A_{0}(y)\right\}_{Dirac} &=\left\{ \pi_{0}(x),\pi_{0}(y)\right\}_{Dirac}
 =\left\{ A_{0}(x),\pi_{0}(y)\right\}_{Dirac}=0 \,,\\
\label{pb2}
\left\{ \tilde{A_{0}}(x),\tilde{\pi_{i}}(y)\right\} &=\left\{ A_{0}(x),\pi_{i}(y)\right\}_{Dirac}=\frac{1}{2m^2}\epsilon_{0kji}V_{k}\partial_j\delta^{3}(x-y) \,,\\
\label{pb3}
\left\{ \tilde{A}_{0}(x),\tilde{A}_{i}(y)\right\} &=\left\{ A_0(x),A_{i}(y)\right\}_{Dirac}=-\frac{1}{m^2}\partial_{i}\delta^{3}(x-y)\,,\\
\label{pb4}
\left\{ \tilde{A}_{i}(x),\tilde{\pi}_{j}(y)\right\} &=\left\{ A_{i}(x),\pi_{j}(y)\right\} _{Dirac}=\delta_{ij}\delta^{3}(x-y) \,.
\end{align}

\ni Here it is important to comment that the Poisson brackets between the gauge invariant variables, Eqs. (\ref{pb1}),  (\ref{pb2}), (\ref{pb3}) and (\ref{pb4}), are the same 
obtained by the Dirac brackets between the original phase space variables. Therefore the GU variables could be an alternative to the usual algorithm that calculates the Dirac brackets in a particular constrained second-class system. For more details see ref.  \cite{jan2}.
Using Eqs. (\ref{gv}), (\ref{gv12}), (\ref{gv13}) and (\ref{gv14}) in the canonical Hamiltonian, Eq. (\ref{hc}), we obtain the gauge invariant Hamiltonian written only in terms of the original
phase space variables
\begin{align}
\label{hcf}
&\tilde{H}=H_c+\int d^3x\frac{\phi_2^2}{2 m^2} \nonumber \\
&=
\int d^{3}x\bigg\{ \frac{1}{2} \pi_i\pi_i+A_{0}\partial_i\pi_i+\frac{1}{4}F_{ij}F_{ij}
+\frac{1}{2}\epsilon_{0kij}\pi_{k}V_{i}A_{j}+\frac{1}{8}\left[\epsilon_{0kij}V_{i}A_{j}\right]^{2}\nonumber\\
&-\frac{1}{4}\epsilon_{0kij}V_{0}A_{k}F_{ij}
+\frac{1}{4}\epsilon_{0kij}V_{k}A_{0}F_{ij}-\frac{m^2}{2}(A_{0}^{2}-A^{2}_i)+\frac{\phi_2^2}{2 m^2}\bigg\} \,.
\end{align}

\ni It is clear that, by construction, we have $ \{ \widetilde{H}, \tilde{\phi}\}=0$.
The invariant Lagrangian can be found by using the functional form of the original Lagrangian, Eq. (\ref{ol}). For more detail
about this procedure see ref.  \cite{jan}. So, the
gauge invariant Lagrangian can be initially written in the following form
\begin{equation}
\label{ol1}
\tilde{L}=\int d^3x-\frac{1}{4}\tilde{F}_{\alpha\nu}\tilde{F}^{\alpha\nu}-\frac{1}{4}\varepsilon{}_{\beta\alpha\rho\varphi}V{}^{\beta}
\tilde{A}^{\alpha}\tilde{F}^{\rho\varphi}
+m^{2}\tilde{A}_{\alpha}\tilde{A}^{\alpha} \,.
\end{equation}

\ni Using Eq. (\ref{gv}) into (\ref{ol1}) we can derive the gauge invariant Lagrangian
\begin{align}
\label{ol2}
&\tilde{L}=\int d^3x \bigg\{ -\frac{1}{2} (\partial_0A_k-\partial_k \tilde{A}_0)(\partial^0A^k-\partial^k \tilde{A}^0) -\frac{1}{4}F_{ij}F^{ij}\nonumber\\
&-\frac{1}{4}\varepsilon_{0kij}V^0 A^k F^{ij}+\frac{1}{4}\varepsilon_{0kij}V{}^k \tilde{A}^0 F^{ij}
+\frac{1}{2}\varepsilon_{0kij}V^k A^i (\partial^j\tilde{A}^0-\partial^0 A^j)\nonumber\\
&+m^{2}\tilde{A}_0\tilde{A}^0+m^{2}A_i A^i \bigg\} \,.
\end{align}

\ni Here it is important to mention that the gauge invariant Lagrangian, Eq. (\ref{ol2}), can not be reduced
 to a covariant form.


\ni Case (ii)  ($\phi_2$ is the gauge symmetry generator)


\ni We will redefine the constraint  $\phi_2$ as 
\begin{eqnarray}
\tilde{\phi}=\frac{\phi_2}{m^2} \,,
\end{eqnarray}

\ni so that 
\begin{eqnarray}
\label{pt}
\Big\{\tilde{\phi}(y),\phi_1(x)\Big\} =\delta^3(x-y).
\end{eqnarray}

\ni Then, we can use the procedure applied in case (i) again. For example, the gauge invariant variable $\tilde{A}_i$
is constructed by the power series of the integral in $\phi_1$
\begin{equation}
\label{gippi0}
\tilde{A}_i(x)=A_i(x)+\int d^{3}yC_{1}(x,y)\phi_1(y)+\int\int d^{3}yd^{3}zC_{2}(x,y,z)\phi_1(y)\phi_1(z)+... \,.
\end{equation}

\ni As a result we have
\begin{eqnarray}
\label{c1pi}
C_{1}(x)=\frac{1}{m^2}\partial_i\delta^3(x-y) \,,
\end{eqnarray}

\ni where we have used Eq. (\ref{pt}) and
\begin{eqnarray}
\label{pif}
\Big\{A_i(x), \tilde{\phi}(y)\Big\}= \frac{1}{m^2}\partial_i\delta^3(x-y) \,.
\end{eqnarray}

\ni We can show that all the correction terms $C_n$ with $n \geq 2$ are zero.
So, using Eqs. (\ref{c1pi}) and (\ref{pif}) into (\ref{gippi0}) we find
\begin{eqnarray}
\label{gv1}
\tilde{A}_{i}=A_{i}+\frac{\partial_{i}\pi_{0}}{m^2} \,. 
\end{eqnarray}

\ni Repeating this same iterative process in order to obtain the other gauge invariant variables we find
\begin{align}
\label{gv2}
\tilde{\pi}_{0}&=0 \,, \\
\label{gv3}
\tilde{\pi}_{i}&=\pi_{i}+\frac{1}{2m^2}\epsilon_{0kim} V_k\partial_{m}\pi_{0} \,,\\
\label{gv4}
\tilde{A}_0&=A_0.
\end{align}

\ni Using Eqs. (\ref{gv1}), (\ref{gv3}) and (\ref{gv4}) into the Hamiltonian in Eq. (6) and noticing that $\tilde{F}_{ij}=F_{ij}$
so that we can derive the gauge invariant Hamiltonian 
\begin{align}
&\tilde{H}=\int dx^{3}\bigg\{ \frac{1}{2}\tilde{\pi}_{k}\tilde{\pi}_k+A_{0}\partial_{k}\tilde{\pi}_{k}+\frac{1}{4}F_{jk}F_{jk}
+\frac{1}{2}\epsilon_{0kij}\tilde{\pi}_k V_{i}\tilde{A}_j\nonumber\\
&+\frac{1}{8}\left[\epsilon_{0kij}V_{i}\tilde{A}_{j}\right]^{2}-\frac{1}{4}\epsilon_{0kij}V_{0}\tilde{A}_{k}F_{ij}+\frac{1}{4}\epsilon_{0kij}V_{k}A_{0}F_{ij}
-\frac{m^2}{2}(A_{0}^{2}-\tilde{A}_{i}^{2})\bigg\}.
\end{align}

\ni The Poisson brackets between the gauge invariant variables are
\begin{align}
\label{g1}
\left\{ \tilde{A_{0}}(x),\tilde{A_{0}}(y)\right\} &=\left\{ \tilde{\pi_{0}}(x),\tilde{\pi_{0}}(y)\right\} =\left\{ \tilde{A_{0}}(x),\tilde{\pi}_{0}(y)\right\} \nonumber\\
=\left\{ A_{0}(x),A_{0}(y)\right\}_{Dirac} &=\left\{ \pi_{0}(x),\pi_{0}(y)\right\}_{Dirac}
=\left\{ A_{0}(x),\pi_{0}(y)\right\}_{Dirac}=0 \,,\\
\label{g2}
\left\{ \tilde{A_{0}}(x),\tilde{\pi_{i}}(y)\right\} &=\left\{ A_{0}(x),\pi_{i}(y)\right\}_{Dirac}=\frac{1}{2m^2}\epsilon_{0kji}V_{k}\partial_j\delta^{3}(x-y) \,,\\
\label{g3}
\left\{ \tilde{A}_{0}(x),\tilde{A}_{i}(y)\right\} &=\left\{ A_{0}(x),A_{i}(y)\right\} _{Dirac}=\frac{1}{m^2}\partial_{i}\delta^{3}(x-y) \,,\\
\label{g4}
\left\{ \tilde{A}_{i}(x),\tilde{\pi}_{j}(y)\right\} &=\left\{ A_{i}(x),\pi_{j}(y)\right\} _{Dirac}=\delta_{ij}\delta^{3}(x-y) \,.
\end{align}

\ni From Eqs. (\ref{g1}), (\ref{g2}), (\ref{g3}) and (\ref{g4}) we can observe that the Poisson brackets calculated between the gauge invariant variables agree with the results obtained by using the Dirac brackets calculated between the original phase space variables.

 The gauge invariant Lagrangian can be obtained by using the same functional form of the original Lagrangian density, Eq. (\ref{ol}),
 where the initial second-class fields are replaced by the gauge invariant variables
\begin{equation}
\label{oli}
L=\int d^3x \left(-\frac{1}{4}\tilde{F}_{\alpha\nu}\tilde{F}^{\alpha\nu}-\frac{1}{4}\varepsilon_{\beta\alpha\rho\varphi}V^\beta \tilde{A}^{\alpha}\tilde{F}^{\rho\varphi}
+m^{2}\tilde{A}_{\alpha}\tilde{A}^{\alpha}\right) \,.
\end{equation}

\ni The GU variable, $\tilde{F}_{\alpha\nu}$, in Eq. (\ref{oli}) is defined as
\begin{eqnarray}
\tilde{F}_{\alpha\nu}=\partial_\alpha\tilde{A}_\nu-\partial_\nu\tilde{A}_\alpha\,.
\end{eqnarray}

\ni We can generalize the gauge invariant variable, Eq. (\ref{gv1}), using the result of the equation of motion, 
$\partial_0\pi_0=-\partial_{k}\pi_{k}-\frac{1}{4}\epsilon_{0kij}V_{k}F_{ij}+m^2 A_{0}=0$, in such way that
we can define $\tilde{A}_\mu$ as
\begin{eqnarray}
\label{egv2}
\tilde{A}_\mu=A_\mu+\frac{1}{m^2} \partial_\mu\pi_0 \,.
\end{eqnarray}

\ni Substituting Eq.(\ref{egv2}) into Eq. (\ref{oli}) and using the fact that $\tilde{F}_{\alpha\nu}=F_{\alpha\nu}$ so the gauge invariant Lagrangian can be reduced to a covariant form
\begin{equation}
\label{olir}
\tilde{L}=\int d^3x\left(-\frac{1}{4}F_{\alpha\nu}F^{\alpha\nu}-\frac{1}{4}\varepsilon_{\beta\alpha\rho\varphi}V^{\beta}\left(A^{\alpha}
+\partial^{\alpha}\theta\right)F^{\rho\varphi}+\frac{m^2}{2}\left(A_{\mu}+\partial_{\mu}\theta\right)\left(A^{\mu}+\partial^{\mu}\theta\right)\right) \,,
\end{equation}

\ni where $\theta$ in Eq. (\ref{olir}) is defined as $\theta\equiv\frac{\pi_{0}}{m^2}$(Stueckelberg
trick  \cite{vyt2}). Here it is important to mention that we can identified the $\theta$ field with the so-called
Stueckelberg scalar whose gauge transformation cancels that of the $A_\mu$ field, therefore making $L$ invariant.
This Lagrangian is the same found by Vytheeswaran \cite{vyt2} with an extra Lorentz symmetry 
violation term. 


\section{Conclusions and final remarks}

In theoretical physics, one of its key improvements was the construction
of gauge invariant systems and its corresponding symmetries, which made the framework of Standard Model possible.  In other words,
gauge field theories have an underlying r\^ole in the discussion of the physical fundamental interactions.
It is also important to mention that, due to 
the presence of symmetries, gauge invariant systems can describe the theoretical models 
in a more complete approach.  

Having said that, in this paper, we have used the improved GU formalism
to discuss gauge invariance and to disclose hidden symmetries that dwell inside the massive CFJ model, which breaks both Lorentz and gauge invariances.
The method provide these results by converting the massive CFJ, which is a second-class system, into a 
first-class one, namely, into a gauge invariant system. This method has the advantage of not introducing
extra variables. Only variables of the original phase space are
utilized. The GU conversion method preserves the degrees of freedom
of the initial system.
This fact can be verified by using the following 
physical degrees of freedom counting formula \cite{ht}, 
$
N_d= N_t-N_{sc}-2  N_{fc} \,\,,
$
where $N_d$ is the number of degrees of freedom, $N_t$ is the total 
number of canonical variables and $N_{sc}$ and $N_{fc}$ are the number 
of second and first-class constraints, respectively. It is
worth to mention that the
obtainment of gauge invariant variables simplifies the derivation of both the
gauge invariant Hamiltonian and  Lagrangian that correspond
to the massive CSF model. 

To sum up the results obtained here, we have obtained two gauge invariant actions dual
to the massive CFJ model, where the St\"uckelberg trick was not necessary, of course.  Besides,
we have shown precisely that the improved GU procedure can lead us to the same
brackets as the ones obtained via Dirac constrained systems method.  
The other positive point is that no extra variables were used in the calculations, as we said before.

\section*{Acknowledgments}

\ni Paulo R. F. Alves and Cleber N. Costa thank CAPES (Coordena\c c\~ao de Ensino Superior) for financial support. Everton M. C. Abreu and Jorge Ananias Neto thank CNPq (Conselho Nacional de Desenvolvimento Cient\' ifico e Tecnol\'ogico), Brazilian scientific support federal agency, for partial financial support, Grants numbers 302155/2015-5 (E.M.C.A.) and 303140/2017-8 (J.A.N.). E.M.C.A. thanks the hospitality of Theoretical Physics Department at Federal University of Rio de Janeiro (UFRJ), where part of this work was carried out.


\begin{thebibliography}{}

\bibitem{cfj} S. M. Carroll, G. B. Field and R. Jackiw, Phys. Rev. D41 (1990) 1231.

\bibitem{ym} Y. M. P. Gomes and P. C. Malta, Phys Rev D94 (2016) 025031

\bibitem{casana2008lorentz}
R. Casana,  M. M. Ferreira Jr and S. J. Rodrigues, Phys. Rev. D 78 (2008) 125013.

\bibitem{alves2016aspectos}
P. R. F. Alves and V. J. V. Otoya, ``Classical Aspects of Electrodynamics of Carroll-Field-Jackiw,"
Multiverso: Revista Eletr{\^o}nica do Campus de Juiz de Fora 1 (2016) 1.

\bibitem{bft}  I. A. Batalin and E. S. Fradkin, Phys. Lett. B 180 (1986) 157;
I. A. Batalin and E. S. Fradkin, Nucl. Phys. B279 (1987) 514;
I.A. Batalin and I.V. Tyutin, Int. J. Mod. Phys. A6 (1991) 3255.

\bibitem{mr} P. Mitra and R. Rajaraman, Ann. Phys. (N.Y.) 203 (1990) 137.

\bibitem{vyt} R. Anishetty and A. S. Vytheeswaran, J. Phys. A26 (1993) 5613; A. S.  Vytheeswaran, Ann. Phys. (N.Y.) 236 (1994) 297.

\bibitem{vyt2} A. S. Vytheeswaran, Int. J. Mod. Phys. A 13 (1998) 765.

\bibitem{dayi} O. F. Dayi, Phys. Lett. B210 (1988) 147.

\bibitem{biz} C. Bizdadea and S. O. Saliu, Europhys. Lett. 32 (1995) 307; Europhys. Lett. 33 (1996) 171(E).

\bibitem{jan} J. Ananias Neto, Braz. J. Phys. 37 (2007) 1106.

\bibitem{mm} M. Monemzadeh,  Aghileh S. Ebrahimi and S. Sramadi, Mod. Phys. Lett. A Vol. 29 (2014) 1450028.

\bibitem{jan2} J. Ananias Neto, ``The gauge unfixing formalism and the solutions of the Dirac bracket
commutators," arXiv: [hep-th] 0904.4711v2.

\bibitem{ht}   M. Henneaux and C. Teitelboim, ``Quantization of gauge systems," Princeton University Press (1992), pag. 29.

\bibitem{Dirac}    P. A. M. Dirac, ``Lectures on Quantum Mechanics", Dover Publications, Mineola, N.Y. (2001).

\bibitem{pcs}   M. I. Park, Y. J. Park, J. Korean Phys. Soc. 31 (1997) 802.






\end{thebibliography}
\end{document}